\documentclass[article,nofootinbib,superscriptaddress,notitlepage]{revtex4-1}
\usepackage{graphicx}
\usepackage{subfigure}
\usepackage{slashed}
\usepackage{simplewick}
\usepackage{feynmf}
\usepackage{color}
\usepackage{rotating}

\newcommand{\PRE}[1]{{#1}} 

\newcommand{\be}{\begin{equation}}
\newcommand{\ee}{\end{equation}}
\newcommand{\bea}{\begin{eqnarray}}
\newcommand{\eea}{\end{eqnarray}}

\newcommand{\eqref}[1]{Eq.~(\ref{#1})}

\def\beq{\begin{eqnarray}}
\def\eeq{\end{eqnarray}}
\def\bea{\begin{eqnarray}}
\def\eea{\end{eqnarray}}

\def\sigmaSI{\sigma_{\rm SI}}
\def\sigmaSD{\sigma_{\rm SD}}

\newcommand{\gsim}{\lower.7ex\hbox{$\;\stackrel{\textstyle>}{\sim}\;$}}
\newcommand{\lsim}{\lower.7ex\hbox{$\;\stackrel{\textstyle<}{\sim}\;$}}

\newcommand{\Dsl}[1]{\slash\hskip -0.20 cm #1}

\usepackage{enumerate}
\include{epsf}

\begin{document}

\preprint{UH511-1210-13}

\title{
\textsc{Matrix element analyses of dark matter scattering and annihilation}
\PRE{\vspace*{0.1in}}
}

\author{Jason Kumar}
\affiliation{\mbox{Department of Physics \& Astronomy, University of
Hawai'i, Honolulu, HI 96822, USA}
}

\author{Danny Marfatia\PRE{\vspace*{.1in}}}
\affiliation{\mbox{Department of Physics \& Astronomy, University of
Kansas, Lawrence, KS 66045, USA}
\PRE{\vspace*{.1in}}
}



\begin{abstract}
\PRE{\vspace*{.1in}}
We provide a compendium of results at the level of matrix elements for a systematic study of dark matter scattering and annihilation.
We identify interactions that yield spin-dependent and spin-independent scattering and specify whether the interactions are
velocity- and/or
momentum-suppressed. We identify the interactions that lead to $s$-wave or $p$-wave annihilation,
and those that are chirality-suppressed. We also list
the interaction structures that can interfere in scattering and annihilation processes. Using these results,
we point out situations in which deviations from the standard lore
are obtained.

\end{abstract}


\maketitle


\section{Introduction}

Recently, several experiments have reported signals that may
be interpreted as hints of dark matter interactions~\cite{DAMA,CoGeNT,CRESST,cdms,PositronExcess,130GeV}.
However, since none of these signals have been recognized as smoking guns for weakly interacting
massive particles (WIMPs) of the Minimal Supersymmetric Standard Model (MSSM), there has been
renewed interest in more general studies of dark matter models.  One
particular area of recent interest is in an effective operator analysis~\cite{EffectiveOperator,Weiler,Freytsis:2010ne,Kumar:2011dr},
where the detailed microscopic physics underlying interactions between
the dark sector and Standard Model (SM) sector are abstracted away, leaving
a description in terms of effective 4-point contact interaction operators.

Thus far, this type of analysis has been carried forward in a somewhat piecemeal
manner.  For example, many analyses assume that dark matter interactions involve
a single contact interaction operator, without accounting for possible effects
arising from interference between multiple operators.  Only initial steps have
been taken toward complementary studies of effective operators using direct, indirect and collider
search strategies.  Although it is well recognized that the effective operator approximation
can break down if the mediating particles are not heavy enough, there has been
little study of the features of the effective operator analysis which are robust.

Our goal is to provide tools needed for a
systematic matrix element study of dark matter interactions with the
Standard Model sector, and results relevant for direct, indirect and
collider searches.  We address the following questions:
\begin{enumerate}
\item{Which interaction structures yield spin-dependent (SD) or spin-independent (SI) scattering?
Are these matrix elements unsuppressed, or suppressed by factors of the relative
velocity or momentum transfer?}
\item{Which interaction structures permit $s$-wave annihilation or $p$-wave
annihilation, and which are chirality-suppressed?}
\item{Which interaction structures can interfere with each other in a scattering process?
Which can interfere in an annihilation process?}
\item{What unique signals arise from interaction structures that are
$CP$-violating?}
\item{How may we distinguish between spin-0, spin-1/2 (Majorana or Dirac)
and spin-1 dark matter by utilizing signals in direct, indirect and collider searches?}
\end{enumerate}

Terms in the scattering matrix element can be suppressed by factors proportional to the
relative velocity, or to the ratio of the momentum transfer to dark matter or nucleus mass.
For cold dark matter, these factors are all very small.
It is common to focus on scattering matrix elements with no
velocity or momentum suppressions, since these terms will typically dominate the scattering
cross section.  However, velocity- or momentum-suppressed terms can dominate if the unsuppressed
terms have very small
coefficients.  We therefore provide a complete treatment of the velocity- and
momentum-suppressed terms as well.

The basic structure of a dark matter-SM interaction can be written in terms
of a dark matter bilinear $\Gamma_X$, and a SM bilinear $\Gamma_f$:
\bea
{\cal O} = \Gamma_X \Gamma_f  F(s,t,u)\,.
\eea
$F$ is a form factor which describes deviations from the structure of a pure contact interaction
(if the form of $F$ is determined for a scattering interaction, it is determined for an annihilation
process by crossing symmetry);
for a contact interaction, $F = \rm{constant}$.
$F$ depends on the details of the particle physics model, including the mass
of the mediating particles, as well as nuclear form factors.  On the other hand, at lowest
dimension, $\Gamma_X$ and $\Gamma_f$ are somewhat more restricted and can be characterized
by their Lorentz structure.  If the interaction structure mediates a process such as $t$-channel
scattering, then the Lorentz structure will be determined by the spin and parity of the mediating
particle which is exchanged.  But if the mediating particle is exchanged in the $u$- or $s$-channel
for a scattering process, then the interaction structure will be more complicated, and can be
determined through use of Fierz transformations.
Our focus will
be on the features that can be determined from knowledge of these bilinears.
In the following, we denote a quark field by $q$, a spin-0 dark matter field
by $\phi$, a spin-1/2 dark matter field by $X$, and a spin-1 dark matter field by $B_\mu$.
A general fermion field (either dark matter or Standard Model) will be represented by
$\psi$.

For simplicity, we focus on interactions only between dark matter and SM
fermions, or with the Higgs.  In Sections II and III, we describe our computation of the scattering
matrix elements and dark matter annihilation matrix elements, respectively.  In section IV we compile our results.  In section V, we conclude
with a discussion of interesting features and deviations from standard lore that arise from the
application of our analysis.

\section{Scattering}

The kinematics of a scattering process in the center of mass frame are determined
by the relative velocity $v$ and the momentum transfer $\overrightarrow{q}$.  In addition
to these kinematic variables, each bilinear can contain terms that are either independent of
spin, or depend on the spin matrix element.  If the spin matrix element is a vector, then it
can be projected on any of three orthogonal axes.  It is useful
to define these three axes as $\hat q = \overrightarrow{q}/ |\overrightarrow{q}|$,
$\hat v^\bot = (\overrightarrow{v}-\overrightarrow{v}\cdot \hat q)/|\overrightarrow{v}-\overrightarrow{v}\cdot \hat q|$
and $\hat \eta = \hat q \times \hat v^\bot$.
In other words, each bilinear will be a sum of terms of the form
\bea
(...) \langle \zeta_{out} | \Gamma | \zeta_{in} \rangle\,,
\eea
where $(...)$ is a function of $\overrightarrow{q}$ and $\overrightarrow{v}$,
$\zeta_{in,out}$ is the spin state of the incoming and outgoing particle, respectively, and
$\Gamma = 1, {S}_{\hat q}\hat q, {S}_{\hat v^\bot} \hat v^\bot , {S}_{\hat \eta} \hat\eta$
(if dark matter is spin-1, then there can also be spin matrix elements which transform as a
symmetric traceless tensor).
Terms with $\Gamma =1$ are spin-independent, while the others are spin-dependent.

The spin and kinematic dependence of these bilinears can most easily be understood from
their transformation properties under rotation and parity.  Each bilinear can
depend on only the incoming and outgoing momenta (which are odd under parity) and the spin
matrix element (which is even under parity).  A bilinear with a single spatial index transforms
as a vector under rotations, while a bilinear with only timelike indices transforms as a scalar.
A bilinear must then consist of a sum of terms in which the momenta and spin are contracted in
such a way as to have the correct rotation and parity transformation properties.

For each bilinear interaction structure, these matrix elements are computed in Appendix~A, and listed
for convenience in Table~\ref{tab:ScatBilinear}.  Also computed
there are squared matrix elements, summed over all initial and final state spins.
By contracting a dark matter bilinear with a SM bilinear, one gets a possible
interaction structure.  From Table~\ref{tab:ScatBilinear}, one can determine the full momentum- and velocity-dependence
of the spin-dependent and spin-independent matrix elements for all such interaction structures.

Also from Table~\ref{tab:ScatBilinear}, we see that there are only a
few Lorentz structures for the Standard Model coupling
such that the nucleon matrix element is momentum- and velocity-independent.
These are $\bar q q$ (SI),  $\bar q \gamma^0 q$ (SI),
$\bar q \gamma^i \gamma^5 q$ (SD), and
$\bar q \sigma^{ij} q $ (SD).

\section{Annihilation}

For the annihilation process, we are guided by the $C$ and $P$ quantum numbers of
the initial and final state.  We assume that both the initial state and
final state consist of a particle and its anti-particle, which may be identical to
the particle.
For a fermion/anti-fermion
state, the transformations under charge conjugation and parity are given by
\bea
C : (-1)^{L+S}
\qquad
P : (-1)^{L+1}\,,
\eea
while for a boson/anti-boson initial states, the transformations are given
by
\bea
C : (-1)^{L+S}
\qquad
P : (-1)^{L}\,.
\eea

The only allowed $s$-wave states are $L=0$, $S=0$ ($J=0$);
$L=0$, $S=1$ ($J=1$); and $L=0$, $S=2$ ($J=2$).  Consequently, we are primarily
interested in initial and final states with $J=0,1,2$.
In Table~\ref{tab:SLJCP}, we list the $C$ and $P$ eigenvalues for
a fermion/anti-fermion state (left) or boson/anti-boson state (right) in
terms of the angular momentum quantum numbers.

\begin{table}[ht]
\begin{center}
\begin{tabular}{|c|c|c|c|c|}
  \hline
  S & L & J & C & P \\
  \hline
  0 & 0 & 0 & + & - \\
  0 & 1 & 1 & - & + \\
  1 & 0 & 1 & - & - \\
  1 & 1 & 0,1,2 & + & + \\
  1 & 2 & 1,2,3 & - & - \\
  1 & 3 & 2,3,4 & + & + \\
  \hline
\end{tabular}
\qquad
\begin{tabular}{|c|c|c|c|c|}
  \hline
  S & L & J & C & P \\
  \hline
  0 & 0 & 0 & + & + \\
  0 & 1 & 1 & - & - \\
  1 & 0 & 1 & - & + \\
  1 & 1 & 0,1,2 & + & - \\
  1 & 2 & 1,2,3 & - & + \\
  2 & 0 & 2 & + & + \\
  2 & 1 & 1,2,3 & - & - \\
  2 & 2 & 0,1,2,3,4 & + & + \\
  2 & 3 & 1,2,3,4,5 & - & - \\
  2 & 4 & 2,3,4,5,6 & + & + \\
  \hline
\end{tabular}
\caption{The $C$ and $P$ transformation properties of a fermion/anti-fermion (left)
or boson/anti-boson (right) state for given $S$, $L$ and $J$ quantum numbers.}
\label{tab:SLJCP}
\end{center}
\end{table}

For any bosonic or fermionic bilinear, the transformation of the bilinear under
rotations determines the total angular momentum of the state that this bilinear
either creates or annihilates.  This information, along with the $C$ and $P$ quantum
numbers of the bilinear, are thus sufficient to determine (from Table~\ref{tab:SLJCP})
the spin and orbital angular momentum of the initial and final state.  The $S$ and
$L$ quantum numbers of the states created (annihilated) by every lowest-dimension
bilinear are listed in Table~\ref{tab:bilinCPJstate}.

\begin{table}[t]
\begin{center}
\begin{tabular}{|c|c|c|c|c|}
  \hline
  bilinear & C & P & J & state \\
  \hline
  $\bar \psi \psi$ & + & + & 0 & $S=1$, $L=1$ \\
  $\imath \bar \psi \gamma^5 \psi$ & + & - & 0 & $S=0$, $L=0$ \\
  $\bar \psi \gamma^0 \psi$ & - & + & 0 & none \\
  $\bar \psi \gamma^i \psi$ & - & - & 1 & $S=1$, $L=0,2$ \\
  $\bar \psi \gamma^0 \gamma^5 \psi$ & + & - & 0 & $S=0$, $L=0$ \\
  $\bar \psi \gamma^i \gamma^5 \psi$ & + & + & 1 & $S=1$, $L=1$ \\
  $\bar \psi \sigma^{0i} \psi$ & - & - & 1 & $S=1$, $L=0,2$ \\
 $\bar \psi \sigma^{ij} \psi$ & - & + & 1 & $S=0$, $L=1$ \\
 $\phi^\dagger \phi$ & + & + & 0 & $S=0$, $L=0$ \\
 $\imath Im (\phi^\dagger  \partial^0 \phi )$ & - & + & 0 & none \\
 $\imath Im (\phi^\dagger  \partial^i \phi )$ & - & - & 1 & $S=0$, $L=1$ \\
 $B^\dagger_\mu B^\mu$ & + & + & 0 & $S=0$, $L=0$; $S=2$, $L=2$ \\
 $\imath Im(B^\dagger_\nu \partial^0 B^\nu )$ & - & + & 0 & none \\
 $\imath Im(B^\dagger_\nu \partial^i B^\nu )$ & - & - & 1 & $S=0$, $L=1$; $S=2$, $L=1,3$ \\
 $\imath (B^{\dagger}_i B_j - B^{\dagger}_j B_i)$ & - & + & 1 & $S=1$, $L=0,2$ \\
 $\imath (B^{\dagger}_i B_0 - B^{\dagger}_0 B_i)$ & - & - & 1 & $S=0$, $L=1$; $S=2$, $L=1,3$ \\
 $\epsilon^{0ijk}B_i \partial_j B_k $ & + & - & 0 & $S=1$, $L=1$ \\
 $-\epsilon^{0ijk}B_0 \partial_j B_k $ & + & + & 1 & $S=2$, $L=2$ \\
 $B^\nu \partial_\nu B_0$ & + & + & 0 & $S=0$, $L=0$; $S=2$, $L=2$ \\
 $B^\nu \partial_\nu B_i$ & + & - & 1 & $S=1$, $L=1$ \\
  \hline
\end{tabular}
\caption{The $C$, $P$ and $J$ quantum numbers of any state
that can be either created or annihilated by the bilinear.  For each possible state, the
$S$ and $L$ quantum numbers are also given.}
\label{tab:bilinCPJstate}
\end{center}
\end{table}

We see that the only dark matter bilinears that can couple to an $s$-wave initial
state are $\imath \bar X \gamma^5 X$, $\bar X \gamma^i X$,
$\bar X \gamma^0 \gamma^5 X$, $\bar X \sigma^{0i} X$,
$\phi^\dagger \phi$, $B^\dagger_{\mu} B^{\mu}$, $\imath (B^{\dagger}_i B_j - B^{\dagger}_j B_i)$
and $B^\nu \partial_\nu B^0$.
Note that the structures $\bar \psi \gamma^0 \psi$,  $\imath Im (\phi^\dagger  \partial^0 \phi )$
and $\imath Im(B^\dagger_\nu \partial^0 B^\nu )$ cannot couple to any state and
cannot contribute to any non-zero annihilation matrix element.

The Standard Model fermion bilinear must be able to produce a final state with
the same $J$ quantum number as the initial state (though the $C$ and $P$
transformations need not be the same, since a general interaction structure
can violate either symmetry).  Thus, the spin and orbital angular momentum of the
final and initial state may be different.

Finally, we address the question of whether or not there is a chirality-suppression
$(\propto m_f^2 / m_X^2)$ of the annihilation matrix element.  This suppression arises
if a SM mass insertion is required to produce a final state with
the appropriate spin.
An outgoing state of SM fermions $\bar f f$ can only be in a $S_z=0$ state if the fermion and
anti-fermion are from different Weyl spinors ($f_L$ and $f_R$).  They are in
an $S_z=\pm 1$ state if the fermion and anti-fermion are from the same Weyl
spinor.  We take the $z$-axis to lie along the direction of motion of
the outgoing fermion and anti-fermion, so $L_z = 0$, and
$J_z = S_z$. (Note that for particles moving along the $z$-axis
it is clear that $L_z =0$, because $Y_{lm}(\theta=0, \phi) \neq 0$ only if $m=0$.)

In Table~\ref{tab:SLJJzFermHel}, we list the possible fermion and anti-fermion helicities for
final states with fixed choices of $S$, $L$,  $J$ and $J_z$.  We assume that the
fermion moves along the $+z$-axis and the anti-fermion along the
$-z$-axis, and that the initial state is written in a basis with angular
momentum projected along the $z$-axis.  In our notation,
$\bar f_L$ is a right-handed anti-fermion, the $CP$-conjugate of $f_L$.
For a SM bilinear to produce one of the listed
final states, it must be able to produce a state with appropriate
$S$, $L$ and $J$ quantum numbers.  The helicities of the produced fermion and
anti-fermion are then determined by the number of Dirac matrices
in the bilinear; a bilinear with an even number of Dirac matrices will
produce a fermion/anti-fermion pair from the same Weyl spinor, while a
bilinear with an odd number of Dirac matrices will produce a pair from
different Weyl spinors.
If a bilinear does not produce a fermion
and anti-fermion of the needed helicities, then there will be a chirality
flip arising from a mass-insertion.

\begin{table}[t]
\begin{center}
\begin{tabular}{|c|c|c|c|c|}
  \hline
  $S$ & $L$ & $J$ & $J_z = S_z$ & fermion helicities \\
  \hline
  0 & 0 & 0 & 0 & $f_L$, $\bar f_R$; $f_R$, $\bar f_L$ \\
  \hline
  1 & 0 & 1 & 1 & $f_R$, $\bar f_R$ \\
  1 & 0 & 1 & 0 & $f_L$, $\bar f_R$; $f_R$, $\bar f_L$ \\
  1 & 0 & 1 & -1 & $f_L$, $\bar f_L$\\
  \hline
  0 & 1 & 1 & 0 & $f_L$, $\bar f_R$; $f_R$, $\bar f_L$ \\
  \hline
  1 & 1 & 0 & 0 & $f_L$, $\bar f_R$; $f_R$, $\bar f_L$ \\
  \hline
  1 & 1 & 1 & 1 & $f_R$, $\bar f_R$\\
  1 & 1 & 1 & 0 & - \\
  1 & 1 & 1 & -1 & $f_L$, $\bar f_L$ \\
  \hline
  1 & 2 & 1 & 1 & $f_R$, $\bar f_R$ \\
  1 & 2 & 1 & 0 & $f_L$, $\bar f_R$; $f_R$, $\bar f_L$ \\
  1 & 2 & 1 & -1 & $f_L$, $\bar f_L$ \\
  \hline
\end{tabular}
\caption{The possible fermion and anti-fermion helicities of a fermion/anti-fermion
state with given $S$, $L$, $J$ and $J_z$ quantum numbers.  It is assumed that the fermion
is travelling on the $+z$-axis, and the anti-fermion is travelling on the $-z$-axis.
$\bar f_{L,R}$ denotes the $CP$-conjugate of $f_{L,R}$ (so, for example, $\bar f_L$ is
a right-handed anti-fermion ).}
\label{tab:SLJJzFermHel}
\end{center}
\end{table}

We can now bring together all of the pieces which contribute to an
understanding of the annihilation matrix element.
The procedure is as follows:
\begin{itemize}
\item{For each interaction structure, we find the $C$ and $P$ transformations
and $J$ quantum number of the dark matter bilinear, and from this
identify the initial state that can couple to this bilinear; $s$-wave
annihilation is only permitted if this state has $L=0$.}
\item{We then determine if the Standard Model bilinear can create a final
state with the same $J$ as the initial state.  If so, the matrix element
for annihilation from the initial state to the appropriate final state
is non-zero.}
\item{We then check if the matrix element has an additional $m_f/m_X$
chirality suppression.  For each $J_z$ projection of the final state, we
find the helicities of the final state fermion and anti-fermion.
If there is no choice of
$J_z$ for which the SM bilinear can produce fermions with the appropriate
helicities, then the annihilation cross section
is suppressed by $m_f^2 / m_X^2$.}
\end{itemize}

In Appendix B, we list the matrix elements arising from fermion/anti-fermion
creation or annihilation, for all choices of interaction structure.  In the
interest of generality, the anti-fermion is not assumed to be the anti-particle
of the fermion, and the two particles are allowed to have different masses.  These
matrix elements can thus be used for the case of dark matter co-annihilation, or if
dark matter annihilates through a flavor-violating process.  The standard case can be
obtained by setting the masses of the two particles to be equal.

\section{Results}

We summarize our results in the following four tables.  In Table~\ref{tab:MainTable}, we
list the dependence of the spin-independent and spin-dependent scattering
matrix elements on $\overrightarrow{S}$, $\overrightarrow{q}$ and $v^{\bot}$.  For
each structure, we indicate whether the momentum or velocity dependence arises from
the dark matter or Standard Model bilinear.  For interactions structures that
yield several matrix element terms with different kinematic dependence,
the kinematic dependence of each term is listed on a separate line.  We also list
if each interaction permits $s$-wave annihilation, and (if so) whether or not
$s$-wave annihilation is chirality-suppressed.  Note that, using Lorentz gauge, one
can rewrite $B_\nu \partial^\nu B_\mu$ as $\partial^{\nu} (B_\nu B_{\mu}) =
\partial^\nu [(1/4)g_{\mu \nu} B^\rho B_\rho + B_\nu B_\mu (sym) ]$, where ``$(sym)$" means symmetric
and traceless in the $\mu \nu$ indices.  Note also that, although operator V8 permits
annihilation from an $s$-wave dark matter initial state, the matrix element nevertheless has
an additional $v^2$-suppression which arises because it depends on the time-like
component of the polarization vector, and thus vanishes in the non-relativistic limit.

\begin{table}[b]
\begin{center}
\begin{tabular}{|c|c|c|c|c|}
  \hline
  Name & Interaction Structure & $\sigmaSI$ suppression & $\sigmaSD$ suppression &$s$-wave?\\
\hline
  F1 & $\bar X X \bar q q$ & 1  & $q^2 v^{\bot 2}$ (SM) & No\\
\hline
  F2 & $\bar X \gamma^5 X \bar q q$ & $q^2 $ (DM) & $q^2 v^{\bot 2}$ (SM); $q^2$ (DM) & Yes \\
\hline
  F3 & $\bar X X \bar q \gamma^5 q$ & 0 & $q^2 $ (SM) & No \\
\hline
  F4 & $\bar X \gamma^5 X \bar q \gamma^5 q$ & 0& $q^2$ (SM); $q^2$ (DM) & Yes \\
\hline
  F5 & $\bar X \gamma^\mu X \bar q \gamma_\mu q$ & 1 & $q^2 v^{\bot 2}$ (SM) & Yes \\
   & (vanishes for Majorana $X$)   & & $q^2$ (SM); $q^2$ or $v^{\bot 2}$ (DM) & \\
\hline
  F6 & $\bar X \gamma^\mu \gamma^5 X \bar q \gamma_\mu q$ & $v^{\bot 2}$ (SM or DM) & $q^2$ (SM) & No \\
\hline
  F7 & $\bar X \gamma^\mu X \bar q \gamma_\mu \gamma^5 q$ & $q^2 v^{\bot 2}$ (SM); $q^2$ (DM) & $v^{\bot 2}$ (SM) & Yes \\
   & (vanishes for Majorana $X$)  & & $v^{\bot 2}$ or $q^2$ (DM)& \\
\hline
  F8 & $\bar X \gamma^\mu \gamma^5 X \bar q \gamma_\mu \gamma^5 q$ & $q^2 v^{\bot 2}$ (SM) & 1 & $\propto m_f^2 / m_X^2$  \\
\hline
  F9 & $\bar X \sigma^{\mu \nu} X \bar q \sigma_{\mu \nu} q$ & $q^2$ (SM); $q^2$ or $v^{\bot 2}$ (DM)  &1 & Yes \\
   &(vanishes for Majorana $X$) & $q^2 v^{\bot 2}$ (SM) & & \\
\hline
  F10 & $\bar X \sigma^{\mu \nu} \gamma^5 X \bar q \sigma_{\mu \nu} q$ & $q^2$ (SM) & $v^{\bot 2}$ (SM) & Yes \\
   & (vanishes for Majorana $X$) & & $q^2$ or $v^{\bot 2}$ (DM) & \\ \hline
\hline
  S1 & $\phi^\dagger \phi \bar q q$ or $\phi^2 \bar q q$ & 1 & $q^2 v^{\bot 2}$ (SM) & Yes \\
\hline
  S2 & $\phi^\dagger \phi \bar q \gamma^5 q$ or $\phi^2 \bar q \gamma^5 q$ & 0 & $q^2$ (SM) & Yes \\
\hline
  S3 & $\phi^\dagger \partial_\mu \phi \bar q \gamma^\mu q$ & 1 & $q^2 v^{\bot 2}$ (SM) & No \\
   & & & $q^2$ (SM); $v^{\bot 2}$ (DM) & \\
\hline
  S4 & $\phi^\dagger \partial_\mu \phi \bar q \gamma^\mu \gamma^5 q$ & 0 & $v^{\bot 2}$ (SM or DM) & No \\ \hline
\hline
  V1 & $B^\dagger_\mu B^\mu \bar q q$ or $B_\mu B^\mu \bar q q$ & 1 & $q^2 v^{\bot 2}$ (SM) & Yes \\
\hline
  V2 & $B^\dagger_\mu B^\mu \bar q \gamma^5 q$ or $B_\mu B^\mu \bar q \gamma^5 q$& 0 & $q^2$ (SM) & Yes \\
\hline
  V3 & $B^\dagger_\nu \partial_\mu B^\nu \bar q \gamma^\mu q$ & 1 & $q^2 v^{\bot 2}$ (SM) & No \\
   & & & $q^2$ (SM); $v^{\bot 2}$ (DM) & \\
\hline
  V4 & $B^\dagger_\nu \partial_\mu B^\nu \bar q \gamma^\mu \gamma^5 q$ & 0 & $v^{\bot 2}$ (SM or DM) & No \\
\hline
  V5 & $(B^\dagger_\mu B_\nu - B^\dagger_\nu B_\mu) \bar q \sigma^{\mu \nu} q$ &
  $q^2 v^{\bot 2}$ (SM) & 1 &  Yes \\
\hline
  V6 & $(B^\dagger_\mu B_\nu - B^\dagger_\nu B_\mu) \bar q \sigma^{\mu \nu} \gamma^5 q$ &
  $q^2$ (SM) & $v^{\bot 2}$ (SM) & Yes \\
\hline
  V7 & $B^\dagger_\nu \partial^\nu B_\mu \bar q \gamma^\mu q$ or $B_\nu \partial^\nu B_\mu \bar q \gamma^\mu q$ & $v^{\bot 2}$ (SM); $q^2$ (DM)
  & $q^2$ (SM); $q^2$ (DM)  & No \\
\hline
  V8 & $B^\dagger_\nu \partial^\nu B_\mu \bar q \gamma^\mu \gamma^5 q$ or $B_\nu \partial^\nu B_\mu \bar q \gamma^\mu \gamma^5 q$ &
  $q^2 v^{\bot 2}$ (SM); $q^2$ (DM) & $q^2$ (DM) & $\propto m_f^2 / m_X^2$ \\
\hline
  V9 & $\epsilon^{\mu \nu \rho \sigma} B^\dagger_\nu \partial_\rho B_\sigma \bar q \gamma_\mu q$ or
  $\epsilon^{\mu \nu \rho \sigma} B_\nu \partial_\rho B_\sigma \bar q \gamma_\mu q$ & $v^{\bot 2}$ (DM or SM) & $q^2$ (SM) & No \\
\hline
  V10 & $\epsilon^{\mu \nu \rho \sigma} B^\dagger_\nu \partial_\rho B_\sigma \bar q \gamma_\mu \gamma^5 q$ or
  $\epsilon^{\mu \nu \rho \sigma} B_\nu \partial_\rho B_\sigma \bar q \gamma_\mu \gamma^5 q$ & $q^2 v^{\bot 2}$ (SM) & 1 & No \\
\hline
\end{tabular}
\caption{The kinematic suppression of the spin-independent and spin-dependent scattering cross sections for all possible interaction structures.
F1-F10 correspond to fermionic dark matter (with F5, F7, F9 and F10 absent for Majorana fermions), S1-S4
correspond to real or complex scalar dark matter,  V1-V10 to
real or complex vector dark matter. Each
suppression is labelled to indicate if it arises from the SM or dark matter
(DM) bilinear.  If a cross section contains several terms with different kinematic
suppressions, each is listed on a separate line.
We also list if $s$-wave annihilation is permitted and
unsuppressed, if it is chirality-suppressed by a factor
$\propto m_f^2 / m_X^2$, or if it is not permitted at all; although the interactions are expressed  in terms of quark fields $q$, by a slight abuse of notation
we allow for annihilation to any pair of SM fermions $\bar{f}f$, each of mass $m_f$. }
\label{tab:MainTable}
\end{center}
\end{table}

\subsection{Interference}

Of course, it is certainly possible for dark matter to couple to Standard Model matter
through a sum of several effective interaction structures.  In that case, it is important
to understand if these operators can interfere.
For the annihilation process, interference can only occur between structures that
annihilate states of the same quantum numbers ($S$, $L$ and $J$) and create states
of the same quantum numbers.
Table~\ref{tab:AnnihInterfere} indicates the interaction structures that can connect initial and final
states for all possible combinations of quantum numbers; interaction operators that
appear on the same line can interfere with one another in
annihilation processes.
In particular, interference between two interactions structures can only occur for $s$-wave annihilation.  It
can be seen from Table~\ref{tab:bilinCPJstate} that if dark matter is spin-0, then
there are no interference terms.

\begin{table}[t]
\begin{center}
\begin{tabular}{|c|c|c|c|c|c|}
  \hline
  $J$ & $S_{init}$ & $L_{init}$ & $S_{final}$ & $L_{final}$ & Interaction structure \\
  \hline
  0 & 0 & 0 & 0 & 0 & $\bar X \gamma^5 X \bar q \gamma^5 q$, $\bar X \gamma^0 \gamma^5 X \bar q \gamma^0 \gamma^5 q$ \\
  0 & 0 & 0 & 1 & 1 & $\bar X \gamma^5 X \bar q q$ \\
  0 & 1 & 1 & 0 & 0 & $\bar X X \bar q \gamma^5 q$ \\
  0 & 1 & 1 & 1 & 1 & $\bar X X \bar q q$ \\
  1 & 0 & 1 & 0 & 1 & $\bar X \sigma^{ij} X \bar q \sigma^{ij} q$ \\
  1 & 0 & 1 & 1 & 0 & $\bar X \sigma^{ij} X \bar q \sigma^{ij} \gamma^5  q$ \\
  1 & 1 & 0 & 0 & 1 & $\bar X \sigma^{ij} \gamma^5 X \bar q \sigma^{ij} q$ \\
  1 & 1 & 0 & 1 & 0 & $\bar X \gamma^i X \bar q \gamma^i q$, $\bar X \sigma^{ij} \gamma^5 X \bar q \sigma^{ij} \gamma^5 q$\\
  1 & 1 & 1 & 1 & 1 & $\bar X \gamma^i \gamma^5 X \bar q \gamma^i \gamma^5 q$ \\
  1 & 1 & 0 & 1 & 1 & $\bar X \gamma^i X \bar q \gamma^i \gamma^5 q$ \\
  1 & 1 & 1 & 1 & 0 & $\bar X \gamma^i \gamma^5 X \bar q \gamma^i q$ \\
    \hline
  0 & 0 & 0 & 0 & 0 &    $B^\dagger_\mu B^\mu \bar q  \gamma^5 q$, $B^\nu \partial_\nu B_0 \bar q  \gamma^0 \gamma^5 q$ \\
  0 & 0 & 0 & 1 & 1 &    $B^\dagger_\mu B^\mu \bar q q$ \\
  0 & 1 & 1 & 0 & 0 & $\epsilon^{0ijk}B_i \partial_j B_k  \bar q  \gamma^0 \gamma^5 q$ \\
  1 & 0 & 1 & 0 & 1 & $\imath (B^\dagger_i B_0 - B^\dagger_i B_0)\bar q \sigma^{0i}  \gamma^5 q$ \\
  1 & 0 & 1 & 1 & 0 & $\imath (B^\dagger_i B_0 - B^\dagger_i B_0)\bar q \sigma^{0i} q$,
  $\imath Im(B^\dagger_\nu \partial_i B^\nu)\bar q \gamma^i q$ \\
  1 & 0 & 1 & 1 & 1 & $\imath Im(B^\dagger_\nu \partial_i B^\nu)\bar q \gamma^i \gamma^5 q$ \\
  1 & 1 & 0 & 0 & 1 & $\imath (B^\dagger_i B_j - B^\dagger_i B_j) \bar q \sigma^{ij}  q$\\
  1 & 1 & 0 & 1 & 0 & $\imath (B^\dagger_i B_j - B^\dagger_i B_j) \bar q \sigma^{ij} \gamma^5 q$\\
  1 & 1 & 1 & 1 & 0 & $B^\nu \partial_\nu B_i \bar q \gamma^i q$ \\
  1 & 1 & 1 & 1 & 1 & $B^\nu \partial_\nu B_i \bar q \gamma^i \gamma^5 q$ \\
  1 & 2 & 2 & 1 & 0 & $\epsilon^{0ijk}B_j \partial_0 B_k \bar q \gamma_i q $\\
  1 & 2 & 2 & 1 & 1 & $\epsilon^{0ijk}B_j \partial_0 B_k \bar q \gamma_i \gamma^5 q $\\
    \hline
\end{tabular}
\end{center}
\caption{The interaction structures that can annihilate an initial state
with quantum numbers $S_{init}$, $L_{init}$ and $J$ and create a final state with
quantum numbers $S_{final}$, $L_{final}$ and $J$.  If two interaction structures are
listed on the same line, then they can interfere in an annihilation process.}
\label{tab:AnnihInterfere}
\end{table}

We now consider interference between different interaction structures in scattering processes.
As we have seen, each of the SM or dark matter bilinears depends on a spin matrix element
which is either spin-independent (1) or depends on a spin projection ($S_{\hat q}$, $S_{\hat v^\bot}$ or $S_{\hat \eta}$,
if the dark matter spin matrix element is a vector).
For the full interaction structure, there are sixteen possible choices of the full spin matrix element.
The four choices that are independent of the quark spin (but may or may not depend on the dark matter spin)
yield spin-independent scattering, while the remaining twelve choices yield spin-dependent scattering.  Two
interaction structures can interfere in a scattering process only if they have the same full spin matrix
element.  Two operators that couple to different spin projections will not interfere as the interference
terms vanish on summing over spins.

We denote the four choices of the spin-independent matrix element by the numbers 1-4, and the twelve choices of the
spin-dependent matrix element by the letters A-L.  We list in Table~\ref{tab:ScatInterfere}, for each interaction structure
for spin-1/2 dark matter, the leading
spin matrix elements.  If an interaction structure contains terms with multiple spin matrix elements, then they
are listed on separate lines.
Note, it is possible for two operators to each interfere with a third, even if they cannot interfere
with each other.  In Table~\ref{tab:ScatInterfereScalar} we list the leading spin matrix elements if dark matter is spin-0,
and in Table~\ref{tab:ScatInterfereVector} we list the spin matrix elements for spin-1 dark matter.  Note that, for
spin-0 dark matter, it is not necessary to list the dark matter spin matrix element, which is always trivial.  Thus,
all interaction structures can interfere for spin-independent scattering of spin-0 dark matter.  There is interference
in the spin-dependent scattering matrix element if two structures couple to the same nucleon spin matrix element.
For spin-1 DM, the interaction structures V7 and V8 couple to a dark matter spin matrix
element that transforms as a traceless symmetric tensor, denoted by $\Pi$.  We represent it by its components
in the orthogonal basis defined by $\hat q$, $\hat v^\bot$ and $\hat \eta$.

\begin{table}[t]
\begin{center}
\begin{tabular}{|c|c|c|c|c|c|c|}
  \hline
   & Interaction Structure & SI ($S_X$-dep.) & SD ($S_X$-dep.) & SD ($S_{SM}$-dep.) & SI Class & SD Class \\
   \hline
  F1 & $\bar X X \bar q q$ & 1 & 1 & $S_{\hat \eta}$ & 1 & C \\
  \hline
  F2 & $\bar X \gamma^5 X \bar q q$ & $S_{\hat q}$ & $S_{\hat q}$
  & $S_{\hat \eta}$ & 2 & F \\
  \hline
  F3 & $\bar X X \bar q \gamma^5 q$ & - & 1 & $S_{\hat q}$ & - & A \\
  \hline
  F4 & $\bar X \gamma^5 X \bar q \gamma^5 q$ & - & $S_{\hat q}$ & $S_{\hat q}$ & - & D \\
  \hline
  F5 & $\bar X \gamma^\mu X \bar q \gamma_\mu q$  & 1 & 1 & $S_{\hat \eta}$  & 1 & C \\
   & (vanishes for Majorana $X$) & & $S_{\hat v^\bot}$ & $S_{\hat v^\bot}$ & & H \\
   & & & $S_{\hat \eta}$ & $S_{\hat \eta}$ & & L \\
   \hline
  F6 & $\bar X \gamma^\mu \gamma^5 X \bar q \gamma_\mu q$  & $S_{\hat v^\bot}$ &
  $S_{\hat \eta}$  & $S_{\hat v^\bot}$  & 3 & K \\
   & & & $S_{\hat v^\bot}$ & $S_{\hat \eta}$ & & I \\
   \hline
  F7 & $\bar X \gamma^\mu X \bar q \gamma_\mu \gamma^5 q$  & $S_{\hat v^\bot}$ & 1 & $S_{\hat v^\bot}$ & 3 & B \\
   & (vanishes for Majorana $X$) & & $S_{\hat v^\bot}$ & $S_{\hat \eta}$ & & I \\
   & & & $S_{\hat \eta}$ & $S_{\hat v^\bot}$ & & K \\
   \hline
  F8 & $\bar X \gamma^\mu \gamma^5 X \bar q \gamma_\mu \gamma^5 q$  & $S_{\hat \eta}$ &
   $S_{\hat q}$  & $S_{\hat q}$ & 4 & D \\
   & & & $S_{\hat v^\bot}$ & $S_{\hat v^\bot}$ & & H \\
   & & & $S_{\hat \eta}$ & $S_{\hat \eta}$ & & L \\
   \hline
  F9 & $\bar X \sigma^{\mu \nu} X \bar q \sigma_{\mu \nu} q$  & 1 , $S_{\hat \eta}$ &
  $S_{\hat q}$ & $S_{\hat q}$ & 1,\,4 & D \\
   & (vanishes for Majorana $X$) & & $S_{\hat v^\bot}$ & $S_{\hat v^\bot}$ & & H \\
   & & & $S_{\hat \eta}$ & $S_{\hat \eta}$ & & L \\
   \hline
  F10 & $\bar X \sigma^{\mu \nu} \gamma^5 X \bar q \sigma_{\mu \nu} q$  & $S_{\hat q}$ &
  1 & $S_{\hat q}$ & 2 & A \\
  & (vanishes for Majorana $X$) & & $S_{\hat q}$ & $S_{\hat \eta}$  & & F \\
  & & & $S_{\hat \eta}$ & $S_{\hat q}$ & & J \\
  \hline
\end{tabular}
\end{center}
\caption{For each interaction structure, we indicate if the scattering matrix element
is independent of the dark matter or SM spin (${1}$), or if it depends on the projection
of the spin on any of three orthogonal axes: the direction of momentum transfer ($\hat q$),
the direction of the relative velocity transverse to the momentum transfer ($\hat v^\bot $),
or the direction perpendicular to $\hat q$ and $\hat v^\bot$  ($\hat \eta =\hat q \times \hat v^\bot$).  If an
interaction structure yields several terms with different spin component dependence, they are
listed on separate lines.  The sixteen possible couplings to dark matter and nucleon spin are
divided into four classes (1-4) that are independent of the nucleon spin, and twelve classes
(A-L) that are nucleon spin-dependent.  For each structure, all of its coupling classes are listed;
if two interaction structures are listed in the same class, then they can interfere.
}
\label{tab:ScatInterfere}
\end{table}

\begin{table}[t]
\begin{center}
\begin{tabular}{|c|c|c|}
  \hline
   & Interaction Structure & SD ($S_{SM}$-dep.) \\
   \hline
   S1 & $\phi^\dagger \phi \bar q q$ or $\phi^2 \bar q q$ & $S_{\hat \eta} $  \\
   \hline
   S2 & $\phi^\dagger \phi \bar q \gamma^5 q$ or $\phi^2 \bar q \gamma^5 q$  & $S_{\hat q} $  \\
   \hline
   S3 & $\phi^\dagger \partial_\mu \phi \bar q \gamma^\mu q$ & $S_{\hat \eta} $  \\
   \hline
   S4 & $\phi^\dagger \partial_\mu \phi \bar q \gamma^\mu \gamma^5 q$ & $S_{\hat v^\bot} $  \\
   \hline
\end{tabular}
\end{center}
\caption{Similar to Table~\ref{tab:ScatInterfere}, but for spin-0 dark matter.
Thus, there is no dependence on the dark matter spin.
All of these structures can interfere for spin-independent scattering.  For spin-dependent
scattering, two interaction structures can interfere if they couple to the same projection
of the nucleon spin.
}
\label{tab:ScatInterfereScalar}
\end{table}

\begin{table}[t]
\begin{center}
\begin{tabular}{|c|c|c|c|c|c|c|}
  \hline
   & Interaction Structure & SI ($S_X$-dep.) & SD ($S_X$-dep.) & SD ($S_{SM}$-dep.) & SI Class & SD Class \\
   \hline
  V1 & $B^\dagger_\mu B^\mu \bar q q$ or $B_\mu B^\mu \bar q q$ & 1  & 1 & $S_{\hat \eta}$ & 1 & C \\
  \hline
  V2 & $B^\dagger_\mu B^\mu \bar q \gamma^5 q$ or $B_\mu B^\mu \bar q \gamma^5 q$ & - & 1 & $S_{\hat q}$ & - & A \\
  \hline
  V3 & $B^\dagger_\nu \partial_\mu B^\nu \bar q \gamma^\mu q$ & 1 & 1 & $S_{\hat \eta}$ & 1 & C \\
  \hline
  V4 & $B^\dagger_\nu \partial_\mu B^\nu \bar q \gamma^\mu \gamma^5 q$ & - & 1 & $S_{\hat v^\bot}$ & 1 & B \\
  \hline
  V5 & $(B^\dagger_\mu B_\nu - B^\dagger_\nu B_\mu)\bar q \sigma^{\mu \nu} q$ & $S_{\hat \eta}$ & $S_{\hat q}$ & $S_{\hat q}$ & 4 & D \\
  & & & $S_{\hat v^\bot}$ & $S_{\hat v^\bot}$ & & H \\
  & & & $S_{\hat \eta}$ & $S_{\hat \eta}$ & & L \\
  \hline
  V6 & $(B^\dagger_\mu B_\nu - B^\dagger_\nu B_\mu)\bar q \sigma^{\mu \nu} \gamma^5 q$ & $S_{\hat q}$ & $S_{\hat q}$ & $S_{\hat \eta}$ & 2 & F \\
  & & & $S_{\hat \eta}$ & $S_{\hat q}$ & & J \\
  \hline
  V7 & $B^\dagger_\nu \partial^\nu B_\mu \bar q \gamma^\mu q$ or $B_\nu \partial^\nu B_\mu \bar q \gamma^\mu q$ & $\Pi_{\hat q \hat v^\bot}$ & $\Pi_{\hat q \hat v^{\bot}}$ & $S_{\hat \eta}$ &  &  \\
  & & & $\Pi_{\hat q \hat \eta }$ & $S_{\hat v^\bot}$ &  &  \\
  \hline
  V8 & $B^\dagger_\nu \partial^\nu B_\mu \bar q \gamma^\mu \gamma^5 q$ or $B_\nu \partial^\nu B_\mu \bar q \gamma^\mu \gamma^5 q$ & $S_{\hat q}$  & $\Pi_{\hat q \hat v}$ & $S_{\hat v}$ & 2 &  \\
  & & $\Pi_{\hat q \hat \eta}$ & $\Pi_{\hat q \hat q}$ & $S_{\hat q}$ &  &  \\
  & & & $\Pi_{\hat q \hat \eta}$ & $S_{\hat \eta}$ &  &  \\
  \hline
  V9 & $\epsilon^{\mu \nu \rho \sigma} B^\dagger_\nu \partial_\rho B_\sigma \bar q \gamma_\mu q$ or
  $\epsilon^{\mu \nu \rho \sigma} B_\nu \partial_\rho B_\sigma \bar q \gamma_\mu q$ & $S_{\hat v^\bot}$  & $S_{\hat v^\bot}$ & $S_{\hat \eta}$ & 3 & I \\
  & & & $S_{\hat \eta}$ & $S_{\hat v^\bot}$ & & K \\
  \hline
  V10 & $\epsilon^{\mu \nu \rho \sigma} B^\dagger_\nu \partial_\rho B_\sigma \bar q \gamma_\mu \gamma^5 q$ or
  $\epsilon^{\mu \nu \rho \sigma} B_\nu \partial_\rho B_\sigma \bar q \gamma_\mu \gamma^5 q$ & $S_{\hat \eta}$ & $S_{\hat q}$ & $S_{\hat q}$ & 4 & D \\
  & & & $S_{\hat v^\bot}$ & $S_{\hat v^\bot}$ & & H \\
  & & & $S_{\hat \eta}$ & $S_{\hat \eta}$ & & L \\  \hline
\end{tabular}
\end{center}
\caption{Similar to Table~\ref{tab:ScatInterfere}, but for spin-1 dark matter.
For structures V7 and V8, the dark matter spin bilinear is a traceless symmetric tensor
represented by $\Pi$ with components
in the orthogonal basis defined by $\hat q$, $\hat v^\bot$ and $\hat \eta$.  Note that V7 and V8 cannot interfere with any
other structure.
}
\label{tab:ScatInterfereVector}
\end{table}

If two interaction structures can interfere, but their matrix elements scale with different powers of
$q$ and $v^{\bot}$, then the interference terms will be small unless one of the structures has a very small
coefficient.  But if two interfering interaction structures are suppressed by the same number of powers
of $q$ and $v^{\bot}$, then the interference terms will be significant as long as the coefficients are
comparable.
In Table~\ref{tab:ScatInterferePower}, we list each interaction structure according to the number of powers of $q$ or $v^\bot$ that appear
in the SI (top) or SD (bottom) matrix element.  Interaction structures that appear within parentheses can interfere.

\begin{table}[t]
\begin{center}
\begin{tabular}{|l|c|c|}
  \hline
 & Powers of $q$ and $v^\bot$ & Interaction structures \\
  \hline
SI &  0 & ($\bar X X \bar q q$, $\bar X \gamma^\mu X \bar q \gamma_\mu q$)  \\
 & 2 &  ($\bar X \gamma^5 X \bar q q$,   $\bar X \sigma^{\mu \nu} \gamma^5 X \bar q \sigma_{\mu \nu} q$),
  $\bar X \gamma^\mu \gamma^5 X \bar q \gamma_\mu q$ \\
 & 4 & ($\bar X \gamma^\mu \gamma^5 X \bar q \gamma_\mu \gamma^5 q$,
  $\bar X \sigma^{\mu \nu} X \bar q \sigma_{\mu \nu} q$) \\
&  6 & $\bar X \gamma^\mu  X \bar q \gamma_\mu \gamma^5 q$\\
  \hline
SD & 0 &  ($\bar X \gamma^\mu \gamma^5 X \bar q \gamma_\mu \gamma^5 q$ , $\bar X \sigma^{\mu \nu}  X \bar q \sigma_{\mu \nu} q$) \\
 & 2 &  ($\bar X X \bar q \gamma^5 q$, $\bar X \sigma^{\mu \nu} \gamma^5  X \bar q \sigma_{\mu \nu} q$),
  ($\bar X \gamma^\mu \gamma^5  X \bar q \gamma_\mu q$, $\bar X \gamma^\mu X \bar q \gamma_\mu \gamma^5 q$) \\
&  4 &  ($\bar X X \bar q  q$, $\bar X \gamma^\mu  X \bar q \gamma_\mu q$), $\bar X \gamma^5  X \bar q \gamma^5 q$\\
&  6 &  $\bar X \gamma^5  X \bar q  q$ \\
  \hline
\end{tabular}
\caption{The number of powers of $q$ and $v^\bot$ that appear in the spin-independent
and spin-dependent scattering cross section for each interaction structure (if dark matter is spin-1/2).  Interaction
structures that are listed together in parentheses can interfere and have the same kinematic suppression.}
\label{tab:ScatInterferePower}
\end{center}
\end{table}

\section{Interesting features and deviations from the standard lore}

These results lead to some interesting observations, including
deviations from the standard lore which arise from consideration of more general models than WIMPs of a
constrained version of the MSSM.  We find:

\begin{enumerate}
\item  The standard lore is that neutralino annihilation to the light Higgs ($XX \rightarrow hh$)
is necessarily $p$-wave suppressed~\cite{Drees:1992am}.  In fact, we see from our analysis that the annihilation of
Majorana fermion dark matter to identical scalars is either $p$-wave suppressed or
suppressed by $CP$-violating phases~\cite{Kumar:2013ira}.  Since the final state consists of identical scalars with
$S=0$, symmetry of the wavefunction requires that $L$ must be even.
If the initial state is $S=0$, $L=0$, $J=0$, $CP$-odd,
then the final state of identical bosons must be $S=0$, $L=0$, $J=0$, $CP$-even, and there
must be $CP$-violation in the annihilation matrix element.  The relevant interaction structure
is then $\bar X \gamma^5 X hh$.
If the initial state is $S=1$, $L=1$, $CP$-even, then
the matrix element is $p$-wave suppressed.

More generally, there are interesting interaction structures which are $CP$-violating
and usually ignored -- so some ``suppressed" annihilation channels can be open
if new physics is $CP$-violating.

\item  The lore is that, if dark matter is a Majorana fermion, then $s$-wave annihilation to
SM fermions is chirality suppressed.  In fact, we find that this is only
true if the term in the dark matter bilinear which annihilates the $s$-wave initial state
couples to the time component of a pseudovector Standard Model bilinear.
For other interaction structures, there need not be any chirality suppression.  From the
point of view of the microscopic theory, these interaction structures can arise from any
new physics which interacts with both Weyl spinors, including sfermion-mixing, heavy fermions,
etc.  Although sfermion-mixing contributions to the matrix element are often assumed
to scale as $m_f/m_X$, this is only true if one makes certain assumptions (such as
minimal flavor violation) about the flavor structure of the theory.

As a concrete example, consider models of isospin-violating dark matter~\cite{IVDM,IVDM1}
that have been entertained in the context of recent signals of low-mass dark matter~\cite{IVDM1}.
The contribution to the spin-independent matrix element from $s$- and $u$-channel squark exchange can be sizable if
left-handed and right-handed squarks mix; squark-mixing for first generation squarks can
therefore contribute to isospin violation.  A consequence of this squark mixing is the presence
of interaction structures other than pseudovector exchange~\cite{Fukushima:2011df}, that can contribute to $s$-wave
annihilation to fermions which is not chirality-suppressed.

\item If the SM fermion bilinear is pseudoscalar ($\bar q \gamma^5 q$), then
the spin-independent scattering matrix element vanishes, including velocity-
or momentum-suppressed terms~\cite{Freytsis:2010ne}.
This can be understood simply from the Lorentz structure of
the interaction; one cannot construct a nucleon matrix element which is invariant under rotations and odd
under parity unless it depends on the nucleon spin.

Interestingly, if the SM fermion bilinear is pseudovector ($\bar q \gamma^\mu \gamma^5 q$) and
the dark matter is spin-0,
then the spin-independent scattering matrix element is zero.  Again, this can be understood
from the Lorentz structure of the interaction.  A spin-independent fermion bilinear matrix element with the
rotation and parity transformation properties needed for a pseudovector coupling must be a vector
proportional to $\overrightarrow{v}^\bot \times \overrightarrow{q}$.  A non-vanishing scalar can only be produced if
this vector is contracted with a dark matter spin polarization, which is not present for spin-0 dark matter.
In fact, if dark matter is spin-1, then the V4 interaction structure will also have exactly vanishing SI matrix
element, because this interaction structure does not depend on the dark matter spin.
Note that these interaction structures involving $\bar q \gamma^\mu \gamma^5 q$ with exactly vanishing
SI-matrix element are all $CP$-violating.

An interesting corollary of this result is that, if it can be shown that SM quarks couple to
dark matter through exchange of a pseudovector, and if a spin-independent scattering cross section can
be measured (even if velocity- or momentum-suppressed), then dark matter cannot be spin-0.  As dark
matter direct detection experiments increase in sensitivity, this result may have useful applications.

\item For spin-1/2 dark matter, only two sets of interaction structures can interfere in an annihilation process, and
both sets annihilate an $L=0$ state.  Only one is relevant for Majorana fermion
dark matter.  If interference occurs for $p$-wave annihilation, then dark matter must be
spin-1.  Spin-0 dark matter does not exhibit interference in annihilation processes.

\item For both SI and SD scattering processes, the interaction structures whose matrix
elements have no velocity- or momentum-suppression can interfere with each other.
But for interaction structures with momentum or velocity-suppressed scattering matrix
elements, interference effects may be small.
For example, the structure $\bar X \gamma^\mu \gamma^5 X \bar q \gamma_\mu q$ yields a
spin-independent scattering cross section which is suppressed by $v^{\bot 2}$, but can
only interfere with interaction structures whose SI matrix element is suppressed by even
more powers of $q$ or $v^\bot$.  For this interaction structure, interference effects will
be small unless it has a very small coefficient.

\item Note that these results do not depend on whether or not the dark matter-SM interaction is short-ranged.
A non-contact interaction can induce additional form factors in the scattering matrix element, but
the kinematic suppressions found here will always be present.  Similarly, although dark matter annihilation
can receive additional suppression if the interaction is non-contact, the results regarding which interaction
structures yield $s$-wave or $p$-wave annihilation are independent of whether or not
the interaction is short-ranged.

\item Most of these interaction structures are dimension 6.  In a collider production process ($pp \rightarrow \bar XX$),
these operators will receive an $\sim (E/\Lambda)^2$ enhancement, where $E$ is the energy scale of
the dark matter production process and $\Lambda$ is the suppression scale of the effective operator.
The dimension 5 interaction structures are
S1, S2, V1, V2, V5 and V6;  these dimension 5 structures will only receive an
$E/ \Lambda$ enhancement.  But if dark matter is a spin-1 particle, then the matrix element can receive
additional $E / m_X$ enhancements from longitudinal polarization tensors.
In particular, an interaction structure will receive an $E / m_X$ enhancement if only
one of the dark matter particles is longitudinally polarized,
while a factor $(E / m_X)^2$ arises if both are.
Scalar dark matter
which interacts through interaction structures S1 and S2 will have suppressed monojet/monophoton
production rates at the LHC~\cite{Kumar:2011dr}.
These features can be used to distinguish between different interaction structures.

\item One can potentially distinguish the dark matter-SM interaction structure by
combining information from indirect, direct and collider search strategies.
As an example, we see that if dark matter is a Majorana fermion, then the only
interaction structure which permits spin-independent scattering without velocity- or
momentum-suppression is $\bar X X \bar q q$ (F1).  But if dark matter is a Dirac fermion, then
there is another interaction structure which permits unsuppressed SI-scattering;
$\bar X \gamma^\mu X \bar q \gamma_\mu q$ (F5).  It is difficult to distinguish these possibilities
with direct detection experiments, but they can be distinguished by the event rates at indirect
detection experiments~\cite{Kumar:2011dr}, since the first operator permits only $p$-wave annihilation (which is
highly suppressed) while the second operator allows annihilation from an $s$-wave state.  However,
if dark matter is a real scalar, then the operator $\phi^2 \bar q q$ (S1) also permits unsuppressed
SI-scattering and $s$-wave annihilation.  This structure can be distinguished from the previous two by monojet and monophoton
searches at the LHC; since this operator is dimension 5, it does not receive as large an energy
enhancement as the other operators~\cite{Kumar:2011dr}.
But if dark matter is spin-1 or is a complex spin-0 particle, then there are other
interaction structures which can yield unsuppressed SI-scattering (S3, V1, V3).
Interaction structures V1 and V3 will yield an LHC production rate with a large enhancement
due to the longitudinal polarization tensors.  This may permit them to be distinguished
from the other interactions structures (and may be distinguished from each other because
V1 allows $s$-wave annihilation, while V3 does not).
However, it is difficult to distinguish structures F1 and S3 without
a more detailed analysis.  Unfortunately, the spin-dependent scattering cross section is not
very useful in distinguishing these two possibilities, since both interaction structures
yield spin-dependent cross sections suppressed by the factor $q^2 v^{\bot 2}$.

\item If the mass of the mediating particle is small compared to the
dark matter mass or the collider production energy scale, then the form factor $F$ will scale
as $E^{-2}$ for a production or annihilation process.  Then, the rate of dark matter annihilation
or production at colliders will be suppressed.  On the other hand, striking signals of the
mediating particle at a collider experiment may then be possible~\cite{An:2012ue}.  Combined studies of indirect detection and
collider production rates can thus provide independent probes of the mass of the particles which mediate
the dark matter interaction.

\item Many directional detection experiments are either operating or under
construction~\cite{Daw:2011wq}.  For such
experiments, the magnitude and direction of $\overrightarrow{q}$ can be measured on an event by
event basis.  With a sufficient number of events at such a detector, one can potentially
distinguish a dependence on $q$ from a dependence on $v^\bot$.  One interesting question has been
the possibility of {\it dark matter astronomy}: the possibility of probing the dark matter velocity
distribution using the event rate of direct detection experiments.  A difficulty is that the event
rate really probes the integral of the velocity distribution.
But since the $v$-dependence of the spin-dependent and spin-independent matrix elements are generally
different, measurements from directional detectors sensitive to SI and SD
scattering can potentially probe two independent moments of the velocity distribution.
This may permit a more detailed study of the velocity distribution of dark matter.
\end{enumerate}

These are immediate and general results which arise from a study of generic dark matter interaction
structures in the formalism which we have described here.  An interesting long-term program for future study
is the use of this formalism to determine the prospects for distinguishing the nature of dark matter
interactions from the many data sets that are becoming available though direct, indirect and
collider experiments.  And if a clear indication of dark matter interactions is discovered, the next
step would be to utilize this formalism to piece together the full dark matter-SM
interaction structure.

\vskip .2in
{\bf Acknowledgments.}
J.~K. (D.~M.) thanks the University of Kansas (Hawaii) for its hospitality while this work was in progress.
We are grateful to A.~Berlin, K.~Fukushima, D.~Hooper, W.-Y. Keung, S.~McDermott, P.~Sandick, P.~Stengel and B.~Thomas for useful discussions.
This research was supported in part by DOE
grants~DE-FG02-04ER41291 and DE-FG02-04ER41308.

\newpage
\appendix
\label{appendix}

\section{Scattering matrix elements}

In Table~\ref{tab:ScatBilinear}, we summarize the bilinear spin matrix elements for an incoming dark matter particle of mass
$m_X$ with momentum
$\overrightarrow{k} = \mu \overrightarrow{v}$ and an outgoing particle with momentum
$\overrightarrow{k'}= \overrightarrow{k}-\overrightarrow{q}$, where
$\mu = m_X m_f / (m_X + m_f)$ is the reduced mass of the dark matter-nucleon system, $\overrightarrow{v}$ is the
relative velocity of the dark matter and the target nucleon, and the $\xi$'s are two-component
spinors; $v^{\bot}$ is defined below.  If dark matter is spin-1, then $\epsilon^\mu$ is its polarization vector.
The spinor bilinears of the Standard Model are related to those
for the dark matter bilinears by $\overrightarrow{q} \rightarrow -\overrightarrow{q}$,
$\overrightarrow{v} \rightarrow -\overrightarrow{v}$, $m_X\rightarrow m_f$, $\xi \rightarrow \zeta$.
We have grouped together terms that have the same Lorentz structure, but including the
possibility of parity violation. The entries of the table are derived below.

\begin{table}[t]
\begin{center}
\begin{tabular}{|c|c|c|}
  \hline
  bilinear &  spin-independent & spin-dependent \\
  \hline
  $\bar \psi \psi$ &   $2m_X \left( \xi'^\dagger  \xi \right)$ &
  $\imath {\mu \over m_X} \epsilon^{ijk} q^i v^{\bot j} \left( \xi'^\dagger \hat S^k \xi \right)$ \\
  $\bar \psi \gamma^5 \psi$ &
  0 & $-2 q^i  \left( \xi'^\dagger \hat{S}^i \xi \right)$ \\
  $\phi^\dagger \phi$  & 1 & 0 \\
  $B^\dagger_\mu B^\mu $  & $\epsilon'^\dagger \cdot \epsilon$ & 0 \\
  \hline
  $\bar \psi \gamma^0 \psi$ &   $2m_X \left( \xi'^\dagger  \xi \right)$ &
  $-\imath {\mu \over m_X} \epsilon^{ijk} q^i v^{\bot j} \left( \xi'^\dagger \hat S^k \xi \right)$ \\
  $\bar \psi \gamma^0 \gamma^5 \psi$ & 0 &   $-4\mu v^{\bot i}  \left( \xi'^\dagger \hat{S}^i \xi \right)$ \\
  $- Im(\phi^\dagger \partial^0 \phi)$  & $m_X$ & 0 \\
  $- Im(B^\dagger_\nu \partial^0 B^\nu)$  & $m_X \epsilon'^\dagger \cdot \epsilon$ & 0 \\
  $\epsilon^{0ijk}B_i \partial_j B_k$ & 0 & $-2\imath \mu \epsilon^{ijk} v^\bot_i \epsilon_j \epsilon'_k $ \\
  \hline
  $\bar \psi \gamma^i \psi$ &    $2\mu v^{\bot i} \left(\xi'^\dagger \xi \right)$ &
   $2 \imath \epsilon^{ijk} q^j  \left(\xi'^\dagger \hat{S}^k \xi \right)$ \\
  $\bar \psi \gamma^i \gamma^5 \psi$ &  $- \imath {\mu \over 2m_X }\epsilon^{ijk}v^{\bot j} q^k \left(\xi'^\dagger \xi \right)$ &
  $-4m_X \left(\xi'^\dagger \hat{S}^i \xi \right)$ \\
  $- Im(\phi^\dagger \partial^i \phi)$  & $\mu v^{\bot i} $ & $0$ \\
  $- Im(B^\dagger_\nu \partial^i B^\nu)$  & $\mu v^{\bot i} \epsilon'^\dagger \cdot \epsilon $ & $0$ \\
  $\epsilon^{0ijk}B_j \partial_0 B_k$  & 0 & $2\imath m_X \epsilon^{ijk} \epsilon_j \epsilon'_k $ \\
  \hline
  $\bar \psi \sigma^{0i} \psi$  &   $q^i \left(\xi'^\dagger \xi \right)$ &
  $4\imath \mu \epsilon^{ijk}v^{\bot ^j} \left(\xi'^\dagger \hat{S}^k \xi \right)$ \\
  $\bar \psi \sigma^{ij} \psi$ &
  $-{\mu \over 2m_X} (q^i v^{\bot j} - q^j v^{\bot i})\left(\xi'^\dagger \xi \right)$ &
  $-4\imath m_X  \epsilon^{ijk} \left(\xi'^\dagger \hat{S}^k \xi \right)$  \\
  $\imath (B^{\dagger i} B^j - B^{\dagger j} B^i)$  & 0 & $\imath (\epsilon'^{\dagger i} \epsilon^j - \epsilon'^{\dagger j} \epsilon^i )$ \\
  $B^i B^j (sym)$ & 0 & $\epsilon'^i \epsilon^j (sym)$ \\
  \hline
\end{tabular}
\caption{For each bilinear for spin-0 ($\phi$), spin-1/2 ($\psi$) and spin-1 ($B_\mu$) particles, we list the spin-independent
and spin-dependent scattering matrix element at leading order. In the last row, ``$(sym)$" means symmetric
and traceless in the $ij$ indices. The spinor bilinears of the Standard Model can be obtained by the substitutions,  $q^i \rightarrow -q^i$,
$v^{\bot i} \rightarrow -v^{\bot i}$, $m_X\rightarrow m_f$, $\xi \rightarrow \zeta$.}
\label{tab:ScatBilinear}
\end{center}
\end{table}

In the center of mass frame, the incoming and outgoing four-momenta of the nucleon, $p$ and $p'$, and the incoming
and outgoing four-momenta of the dark matter, $k$ and $k'$, to first order in the three-momenta, are
\bea
p &=& \bigg(\sqrt{m_f^2 +\overrightarrow{p}^2 }, \overrightarrow{p}\bigg) \sim \big(m_f , \overrightarrow{p}\big)\,,
\nonumber\\
p' &=& \bigg(\sqrt{m_f^2 + (\overrightarrow{p} +\overrightarrow{q})^2}, \overrightarrow{p}+\overrightarrow{q}\bigg)
\sim \big(m_f, \overrightarrow{p} + \overrightarrow{q} \big)\,,
\nonumber\\
k &=& \bigg(\sqrt{m_X^2 + \overrightarrow{k}^2},\overrightarrow{k}\bigg) \sim \big(m_X, \overrightarrow{k}\big)\,,
\nonumber\\
k' &=& \bigg(\sqrt{m_X^2 + (\overrightarrow{k}-\overrightarrow{q})^2},\overrightarrow{k}-\overrightarrow{q}\bigg)
\sim \big(m_X, \overrightarrow{k}-\overrightarrow{q}\big)\,.
\eea
It is useful to define $v^\bot$ via
\bea
2\overrightarrow{k}-\overrightarrow{q} &=& -(2\overrightarrow{p}+\overrightarrow{q})
=2\mu v^\bot
\equiv 2\mu {(\overrightarrow{q}\times \overrightarrow{v})\times \overrightarrow{q} \over |\overrightarrow{q}|^2}\,.
\eea

We compute spinor bilinear matrix elements of the form
\bea
\bar X \Gamma X\,,
\eea
where $\Gamma$ is a Dirac structure.  Where helpful, we also compute the square of the
matrix element, summed over initial and final spins.

We write our spinors as
\bea
u(p) &=& \bigg({p \cdot \sigma + m_f \over \sqrt{2(p^0 + m_f)}} \zeta \ \ \ \ \ \  {p \cdot \bar \sigma + m_f \over \sqrt{2(p^0 + m_f)}} \zeta \bigg)^T\,,
\nonumber\\
u(k) &=& \bigg({k \cdot \sigma + m_X \over \sqrt{2(k^0 + m_X)}} \xi \ \ \ \ \ \  {k \cdot \bar \sigma + m_X \over \sqrt{2(k^0 + m_X)}} \xi \bigg)^T\,.
\eea

\subsection{Scalar}
For a scalar Lorentz structure ($\Gamma =1$), the matrix element is
invariant under rotations and even under parity.  It can therefore only contain terms
that are either constant, or proportional to $\overrightarrow{p'} \cdot \overrightarrow{p}$
or $\overrightarrow{p'} \times \overrightarrow{p} \cdot \overrightarrow{S}$.

We get
\bea
{\cal M}_s &=& \bar u(p') u(p)
\nonumber\\
&=& {1 \over \sqrt{(p'^0+m)(p^0+m)}} \zeta'^\dagger \left[m^2 + p' \cdot p
-  \imath \epsilon ^{ijk} p'^i p^j \sigma^k +m (p'^0 + p^0)
\right] \zeta
\nonumber\\
&\sim& 2m \left( \zeta'^\dagger  \zeta \right) - {\imath \over m} \epsilon^{ijk} p'^i p^j \left( \zeta'^\dagger \hat S^k \zeta \right)\,.
\eea
The dominant term is spin-independent, and the spin-dependent term is suppressed.

Then, the Standard Model matrix element is
\bea
{\cal M}_{s(SM)}
&\sim& 2m_f \left( \zeta'^\dagger  \zeta \right) - {\imath \over m_f} \epsilon^{ijk} q^i p^j \left( \zeta'^\dagger \hat S^k \zeta \right)
\nonumber\\
&\sim&
2m_f \left( \zeta'^\dagger  \zeta \right) + \imath {\mu \over m_f} \epsilon^{ijk} q^i v^{\bot j} \left( \zeta'^\dagger \hat S^k \zeta \right)\,,
\eea
and the dark matter matrix element is
\bea
{\cal M}_{s(X)}
&\sim& 2m_X \left( \xi'^\dagger  \xi \right) + {\imath \over m_X} \epsilon^{ijk} q^i k^j \left( \xi'^\dagger \hat S^k \xi \right)
\nonumber\\
&\sim&
2m_X \left( \xi'^\dagger  \xi \right) + \imath {\mu \over m_X} \epsilon^{ijk} q^i v^{\bot j} \left( \xi'^\dagger \hat S^k \xi \right)\,.
\eea

If dark matter is spin-0 and the bilinear has a scalar Lorentz structure ($\phi^\dagger \phi$), then the matrix element
is
\bea
{\cal M}_{s(X)}^{spin-0} &=& 1\,.
\eea
Similarly, if dark matter is a real or complex spin-1 and the bilinear has a scalar Lorentz structure ($B^\dagger_\mu B^\mu$), then
the matrix element is
\bea
{\cal M}_{s(X)}^{spin-1} &=& \epsilon'^\dagger \cdot \epsilon \,,
\eea
where $\epsilon$ and $\epsilon'$ are polarization vectors.

\subsection{Pseudoscalar}
For a bilinear with pseudoscalar Lorentz structure, the matrix element is invariant under
rotations, but odd under parity.  It must then be proportional to either $\overrightarrow{p}\cdot \overrightarrow{S}$ or
$\overrightarrow{p'}\cdot \overrightarrow{S}$.
The matrix element is given by
\bea
{\cal M}_{ps} &=& \bar u(p') \gamma^5 u(p)
\nonumber\\
&=& {1 \over  \sqrt{(p'^0+m)(p^0+m)}} \zeta'^\dagger \left[
(p^0 +m) (\overrightarrow{p'} \cdot \overrightarrow{\sigma}) - (p'^0 +m) (\overrightarrow{p} \cdot \overrightarrow{\sigma})
\right] \zeta
\nonumber\\
&\sim& (\overrightarrow{p'} - \overrightarrow{p}) \cdot \zeta'^\dagger \overrightarrow{\sigma} \zeta
\sim 2(p' - p)^i \left( \zeta'^\dagger \hat{S}^i \zeta \right)\,,
\eea
which gives
\bea
{\cal M}_{ps(SM)} &=& \bar u(p') \gamma^5 u(p) \sim 2 q^i  \left( \zeta'^\dagger \hat{S}^i \zeta \right)\,,
\nonumber\\
{\cal M}_{ps(X)} &=& \bar u(k') \gamma^5 u(k) \sim -2 q^i  \left( \xi'^\dagger \hat{S}^i \xi \right)\,.
\eea
This structure is spin-dependent and velocity-dependent; interestingly, there is no spin-independent term at all.

\subsection{Vector}
For a bilinear with vector Lorentz structure, the time-like component is invariant under
rotations and parity.  So it can either be a constant, or be proportional to $\overrightarrow{p'}
\times \overrightarrow{p} \times \overrightarrow{S}$.  The space-like
components must rotate as a vector, but be odd under parity.  They may then contain terms that are
proportional to either $\overrightarrow{p'}$, $\overrightarrow{p}$, $\overrightarrow{p'} \times \overrightarrow{S}$
or $\overrightarrow{p} \times \overrightarrow{S}$.

We get
\bea
{\cal M}_v^0 &=& \bar u (p') \gamma^0 u(p)
\nonumber\\
&=& {1 \over  \sqrt{(p'^0+m)(p^0+m)}} \zeta'^\dagger \left[
(p'^0 +m )(p^0 +m) + \overrightarrow{p'}\cdot \overrightarrow{p}
+  \imath \epsilon ^{ijk} p'^i p^j \sigma^k
\right] \zeta
\nonumber\\
&\sim& 2m \left( \zeta'^\dagger \zeta \right) +{\imath \over m} \epsilon^{ijk} p'^i p^j \left( \zeta'^\dagger \hat S^k \zeta \right)\,,
\eea
\bea
{\cal M}_v^i &=& \bar u (p') \gamma^i u(p)
\nonumber\\
&=& {1 \over  \sqrt{(p'^0+m)(p^0+m)}} \zeta'^\dagger \left[(m +p^0)p'^i + (m + p'_0) p^i
+ \imath (m+p'^0) \epsilon^{ijk} p^j  \sigma^k
 -\imath (m+p^0) \epsilon^{ijk} p'^j \sigma^k  \right] \zeta
\nonumber\\
&\sim& (p' + p)^i \left(\zeta'^\dagger \zeta \right) +2 \imath \epsilon^{ijk} (p-p')^j  \left(\zeta'^\dagger \hat{S}^k \zeta \right)\,.
\eea
The leading term of this matrix element is spin-independent, but there are also momentum-suppressed
spin-dependent terms.

We thus find, for the SM matrix elements,
\bea
{\cal M}_{v(SM)}^0 &\sim& 2m_f \left( \zeta'^\dagger \zeta \right)
+{\imath \over m_f} \epsilon^{ijk} q^i p^j \left( \zeta'^\dagger \hat S^k \zeta \right)
\nonumber\\
&\sim& 2m_f \left( \zeta'^\dagger \zeta \right)
-\imath {\mu \over m_f} \epsilon^{ijk} q^i v^{\bot j} \left( \zeta'^\dagger \hat S^k \zeta \right)\,,
\nonumber\\
{\cal M}_{v(SM)}^i  &\sim& (2p+q)^i \left(\zeta'^\dagger \zeta \right) -2 \imath \epsilon^{ijk} q^j  \left(\zeta'^\dagger \hat{S}^k \zeta \right)
\nonumber\\
&\sim& -2\mu v^{\bot i} \left(\zeta'^\dagger \zeta \right) -2 \imath \epsilon^{ijk} q^j  \left(\zeta'^\dagger \hat{S}^k \zeta \right)\,,
\eea
and for the dark matter matrix elements,
\bea
{\cal M}_{v(X)}^0 &\sim& 2m_X \left( \xi'^\dagger \xi \right) -{\imath \over m_X} \epsilon^{ijk} q^i k^j \left( \xi'^\dagger \hat S^k \xi \right)
\nonumber\\
&\sim& 2m_X \left( \xi'^\dagger \xi \right)
-\imath {\mu \over m_X} \epsilon^{ijk} q^i v^{\bot j} \left( \xi'^\dagger \hat S^k \xi \right)\,,
\nonumber\\
{\cal M}_{v(X)}^i &\sim& (2k -q)^i \left(\xi'^\dagger \xi \right) +2 \imath \epsilon^{ijk} q^j  \left(\xi'^\dagger \hat{S}^k \xi \right)
\nonumber\\
&\sim& 2\mu v^{\bot i} \left(\xi'^\dagger \xi \right) +2 \imath \epsilon^{ijk} q^j  \left(\xi'^\dagger \hat{S}^k \xi \right)\,.
\eea

The squared matrix elements can be written as the tensors,
\bea
T_v^{00} &=& 2m (2m+E_R)\sum_{spins} \zeta^\dagger \zeta  =8m^2 +2\overrightarrow{q}^2\,,
\nonumber\\
T_v^{0i} &=& -m \sum_{spins} \zeta^\dagger [(\sigma^\alpha \sigma^i
- \bar \sigma^\alpha \sigma^i) p'_{\alpha} ] \zeta
=-2m q_k \sum_{spins} \zeta^\dagger \sigma^k \sigma^i \zeta
= -2m q^i \sum_{spins} \zeta^\dagger \zeta =-4mq^i\,,
\nonumber\\
T_v^{ij}&=& m \sum_{spins} \zeta^\dagger [(\sigma^i \sigma^\alpha \sigma^j
+\sigma^i \bar \sigma^\alpha \sigma^j) p'_{\alpha} -2m \sigma^i \sigma^j] \zeta
= 2m \sum_{spins} \zeta^\dagger [\sigma^i \sigma^j (m+E_R) -m \sigma^i \sigma^j] \zeta
=2\overrightarrow{q}^2 \delta^{ij}\,.
\eea
So $T_v^{00} \sim 8m^2$, with all other components momentum-suppressed.

If dark matter is a spin-0 particle and the bilinear has a vector Lorentz structure ($- Im (\phi^\dagger \partial^\mu \phi)$),
then
\bea
{\cal M}_v^{(spin-0)\mu} &=& {1 \over 2} (p+p')^\mu\,.
\eea
Similarly, if dark matter is a complex spin-1 particle and the bilinear has a vector Lorentz structure
($- Im (B^\dagger_\nu \partial^\mu B^\nu)$), then
\bea
{\cal M}_v^{(spin-1)\mu} &=& {1 \over 2}(p+p')^\mu \epsilon'^\dagger \cdot \epsilon\,.
\eea
Thus,
\bea
{\cal M}_{v(X)}^{(spin-0)0} &\sim& m_X\,,
\nonumber\\
{\cal M}_{v(X)}^{(spin-0)i} &\sim& \mu v^{\bot i}\,,
\nonumber\\
{\cal M}_{v(X)}^{(spin-1)0} &\sim& m_X \epsilon'^\dagger \cdot \epsilon \,,
\nonumber\\
{\cal M}_{v(X)}^{(spin-1)i} &\sim& \mu v^{\bot i} \epsilon'^\dagger \cdot \epsilon \,.
\eea

\subsection{Pseudovector}
For a bilinear with a pseudovector Lorentz structure, the time-component is rotation-invariant and odd under
parity.  It must therefore be a sum of terms that are proportional to either $\overrightarrow{p}\cdot \overrightarrow{S}$ or
$\overrightarrow{p'} \cdot \overrightarrow{S}$.  The spacelike components must
rotate like a vector, but be even under parity.  They must then be a sum of terms proportional to
either $\overrightarrow{S}$ or $\overrightarrow{p'} \times \overrightarrow{p}$.
We find
\bea
{\cal M}_{pv}^0 &=& \bar u (p') \gamma^0 \gamma^5 u(p)
\nonumber\\
&=& -{1 \over \sqrt{(p'^0+m)(p^0+m)}} \zeta'^\dagger \left[
(p'^0 +m) (\overrightarrow{p} \cdot \overrightarrow{\sigma}) + (p^0 +m) (\overrightarrow{p'} \cdot \overrightarrow{\sigma})
\right] \zeta
\nonumber\\
&\sim& - 2(p' +p)^i \left( \zeta'^\dagger \hat{S}^i  \zeta \right)\,,
\eea
\bea
{\cal M}_{pv}^i &=& \bar u (p') \gamma^i \gamma^5 u(p)
\nonumber\\
&=& -{1 \over  \sqrt{(p'^0+m)(p^0+m)}} \zeta'^\dagger \left[
(p' \cdot p + m(p^0+p'^0)+m^2)\sigma^i
+p^i p'^j \sigma^j + p'^i p^j \sigma^j + \imath \epsilon^{ijk}p^j p'^k
\right] \zeta
\nonumber\\
&\sim& -4m \left(\zeta'^\dagger \hat{S}^i \zeta \right)
+ {\imath \over 2m }\epsilon^{ijk}p^j p'^k \left(\zeta'^\dagger \zeta \right)\,.
\eea
Thus, the leading term of the pseudovector structure is spin-dependent.  Interestingly, while the timelike component has no spin-independent
contribution, the
spacelike components have suppressed spin-independent terms.

Then, the SM matrix elements are
\bea
{\cal M}_{pv(SM)}^0 &\sim& - 2(2p+q)^i \left( \zeta'^\dagger \hat{S}^i  \zeta \right)
\nonumber\\
&\sim&  4\mu v^{\bot i} \left( \zeta'^\dagger \hat{S}^i  \zeta \right)\,,
\nonumber\\
{\cal M}_{pv(SM)}^i &\sim& -4m_f \left(\zeta'^\dagger \hat{S}^i \zeta \right)
+ {\imath \over 2m_X }\epsilon^{ijk}p^j q^k \left(\zeta'^\dagger \zeta \right)
\nonumber\\
&\sim& -4m_f \left(\zeta'^\dagger \hat{S}^i \zeta \right)
-\imath {\mu \over 2m_X }\epsilon^{ijk} v^{\bot j} q^k \left(\zeta'^\dagger \zeta \right)\,,
\eea
and the dark matter matrix elements are
\bea
{\cal M}_{pv(X)}^0 &\sim& - 2(2k-q)^i \left( \xi'^\dagger \hat{S}^i  \xi \right)
\nonumber\\
&\sim& - 4 \mu v^{\bot i} \left( \xi'^\dagger \hat{S}^i  \xi \right)\,,
\nonumber\\
{\cal M}_{pv(X)}^i &\sim& -4m_X \left(\xi'^\dagger \hat{S}^i \xi \right)
- {\imath \over 2m_X }\epsilon^{ijk}k^j q'^k \left(\xi'^\dagger \xi \right)
\nonumber\\
&\sim& -4m_X \left(\xi'^\dagger \hat{S}^i \xi \right)
- \imath {\mu \over 2m_X }\epsilon^{ijk}v^{\bot j} q'^k \left(\xi'^\dagger \xi \right)\,.
\eea

We can write the squared matrix elements as tensors:
\bea
T_{pv}^{00} &=& m \sum_{spins} \zeta^\dagger [( \sigma^\alpha +\bar \sigma^\alpha ) p'_{\alpha} -2m ] \zeta
=2mE_R \sum_{spins} \zeta^\dagger \zeta = 2\overrightarrow{q}^2\,,
\nonumber\\
T_{pv}^{0i} &=& -m \sum_{spins} \zeta^\dagger [(\sigma^\alpha \sigma^i
- \bar \sigma^\alpha \sigma^i) p'_{\alpha} ] \zeta
=-2m q_k \sum_{spins} \zeta^\dagger \sigma^k \sigma^i \zeta
= -2m q^i \sum_{spins} \zeta^\dagger \zeta =-4mq^i\,,
\nonumber\\
T_{pv}^{ij}&=& m \sum_{spins} \zeta^\dagger [(\sigma^i \sigma^\alpha \sigma^j
+\sigma^i \bar \sigma^\alpha \sigma^j) p'_{\alpha} +2m \sigma^i \sigma^j] \zeta
\nonumber\\
&=& 2m \sum_{spins} \zeta^\dagger [\sigma^i \sigma^j (m+E_R) +m \sigma^i \sigma^j] \zeta
=(4m^2 +\overrightarrow{q}^2)\delta^{ij} \sum_{spins} \zeta^\dagger \zeta
\nonumber\\
&=& {16\over 3} J(J+1)(2J+1) \left( m^2 +{\overrightarrow{q}^2\over4} \right) \delta^{ij}\,.
\eea
Thus, $T_{pv}^{i j} = {16\over 3} J(J+1)(2J+1) m^2 \delta^{ij}$ is the dominant component, with all the others momentum-suppressed.

If dark matter is spin-1, then there is one other possible structure, $\epsilon^{\mu \nu \rho \sigma} B_\nu \partial_\rho B_\sigma$, which gives
\bea
{\cal M}_{pv}^{spin-1} &=& \imath \epsilon^{\mu \nu \rho \sigma} (p+p')_\rho \epsilon_\nu \epsilon'_\sigma\,,
\nonumber\\
\eea
so that
\bea
{\cal M}_{pv(X)}^{(spin-1)0} &\sim& -2\imath \mu \epsilon^{ijk} v^\bot_i \epsilon_j \epsilon'_k\,,
\nonumber\\
{\cal M}_{pv(X)}^{(spin-1)i} &\sim& 2\imath m_X \epsilon^{ijk} \epsilon_j \epsilon'_k\,.
\eea

\subsection{Tensor}
Under parity, $\bar \psi \sigma^{\mu \nu} \psi$ transforms with the sign
$(-1)^\mu (-1)^\nu$, where $(-1)^\mu\equiv 1$ for $\mu=0$ and $(-1)^\mu\equiv -1$ for $\mu=1,2,3$.  The structure $\bar \psi \sigma^{0i} \psi$ thus
rotates as a vector, but is odd under parity.  It thus therefore be a sum of terms
proportional to
$\overrightarrow{p'}$, $\overrightarrow{p}$, $\overrightarrow{p'} \times \overrightarrow{S}$
or $\overrightarrow{p} \times \overrightarrow{S}$.  Similarly, the structure
$\bar \psi \sigma^{ij} \psi$ should transform under rotations as a tensor, and
be invariant under parity.  So it should contain terms which are proportional to
$p'^i p^j$ or $S^i S^j$.
We find
\bea
{\cal M}_t^{0i} &=& \bar u (p') \gamma^0 \gamma^i u(p)
\nonumber\\
&=& {1 \over  \sqrt{(p'^0+m)(p^0+m)}} \zeta'^\dagger \left[
-(p^0 +m)(p'^i -\imath \epsilon^{ijk} p'^j \sigma^k)  + (p'^0 +m)(p^i +\imath \epsilon^{ijk} p^j \sigma^k)
\right] \zeta
\nonumber\\
&\sim&
-(p'-p)^i \left(\zeta'^\dagger \zeta \right)
+2\imath \epsilon^{ijk}(p' + p)^j \left(\zeta'^\dagger \hat{S}^k \zeta \right)\,,
\eea
\bea
{\cal M}_t^{ij} &=& \bar u (p') \gamma^{[i} \gamma^{j]} u(p)
\nonumber\\
&=& -{\imath \epsilon^{ijk} \over  \sqrt{(p'^0+m)(p^0+m)}} \zeta'^\dagger \left[
[(p'^0 +m)(p^0+m)-\overrightarrow{p'} \cdot \overrightarrow{p} ]\sigma^k
- p^k p'^l  \sigma^l + \imath \epsilon^{kml} p'^l p^m
+ p'^k p^m \sigma^m
\right] \zeta
\nonumber\\
&\sim& -4\imath m  \epsilon^{ijk} \left(\zeta'^\dagger \hat{S}^k \zeta \right)
+{1\over 2m} (p'^i p^j - p'^j p^i)\left(\zeta'^\dagger \zeta \right)\,.
\eea
This structure is spin-dependent, and also has a momentum-suppressed
spin-independent term.

For the SM matrix elements, we get
\bea
{\cal M}_{t(SM)}^{0i} &=& -q^i \left(\zeta'^\dagger \zeta \right)
+2\imath \epsilon^{ijk} (2p+q)^j \left(\zeta'^\dagger \hat{S}^k \zeta \right)
\nonumber\\
&=& -q^i \left(\zeta'^\dagger \zeta \right)
-4\imath \mu \epsilon^{ijk} v^{\bot j} \left(\zeta'^\dagger \hat{S}^k \zeta \right)\,,
\nonumber\\
{\cal M}_{t(SM)}^{ij} &=& -4\imath m_f  \epsilon^{ijk} \left(\zeta'^\dagger \hat{S}^k \zeta \right)
+{1\over 2m_f} (q^i p^j - q^j p^i)\left(\zeta'^\dagger \zeta \right)
\nonumber\\
&=& -4\imath m_f  \epsilon^{ijk} \left(\zeta'^\dagger \hat{S}^k \zeta \right)
-{\mu \over 2m_f} (q^i v^{\bot j} - q^j v^{\bot i})\left(\zeta'^\dagger \zeta \right)\,,
\eea
and for the dark matter matrix elements, we get
\bea
{\cal M}_{t(X)}^{0i} &=& q^i \left(\xi'^\dagger \xi \right)
+2\imath \epsilon^{ijk}(2k-q)^j \left(\xi'^\dagger \hat{S}^k \xi \right)
\nonumber\\
&=& q^i \left(\xi'^\dagger \xi \right)
+4\imath \mu \epsilon^{ijk}v^{\bot ^j} \left(\xi'^\dagger \hat{S}^k \xi \right)\,,
\nonumber\\
{\cal M}_{t(X)}^{ij} &=& -4\imath m_X  \epsilon^{ijk} \left(\xi'^\dagger \hat{S}^k \xi \right)
-{1\over 2m_X} (q^i k^j - q^j k^i)\left(\xi'^\dagger \xi \right)
\nonumber\\
&=& -4\imath m_X  \epsilon^{ijk} \left(\xi'^\dagger \hat{S}^k \xi \right)
-{\mu \over 2m_X} (q^i v^{\bot j} - q^j v^{\bot i})\left(\xi'^\dagger \xi \right)
\nonumber\\
&=& -4\imath m_X  \epsilon^{ijk} \left(\xi'^\dagger \hat{S}^k \xi \right)\,.
\eea

The dominant squared matrix elements are
\bea
T_t^{\mu \nu \rho \sigma} &=& \sum_{spins} \bar u(p)\sigma^{\mu \nu} (\Dsl p' +m)
\sigma^{\rho \sigma} u(p)
= m\sum_{spins} \bar u(p) \sigma^{\mu \nu}(1 +\gamma^0 ) \sigma^{\rho \sigma} u(p)\,,
\nonumber\\
T_t^{\mu \nu kl}
&=& m\sum_{spins} \bar u(p) \sigma^{\mu \nu} \sigma^{kl}(1 +\gamma^0 )u(p)\,,
\nonumber\\
T_t^{ij kl}
&=& {16J(J+1)(2J+1) \over 3}m^2 (g^{ik}g^{jl} - g^{il}g^{jk})\,,
\eea
with all other components momentum-suppressed.

If dark matter is a complex spin-1 particle with tensor Lorentz structure ($\imath (B^\dagger_\mu B_\nu - B^\dagger_\nu B_\mu)$)
then the matrix element is
\bea
{\cal M}_{t(X)}^{(cpx.~spin-1)\mu \nu} &=& \imath (\epsilon'^{*\mu} \epsilon^\nu - \epsilon'^{*\nu} \epsilon^\mu )\,.
\eea
For a real spin-1 particle with tensor Lorentz structure ($B^\mu B^\nu (sym)$, where ``$(sym)$" means symmetric
and traceless in the $\mu \nu$ indices), the matrix element is
\bea
{\cal M}_{t(X)}^{(real~spin-1)\mu \nu} &=& \epsilon'^\mu \epsilon^\nu (sym)\,.
\eea

\section{Annihilation matrix elements}

We begin by listing the (exact) spinor bilinears for a dark matter creation or annihilation process.
For the sake of generality, we allow the particles to have different masses.
In terms of two-component spinors $\xi_i$, the Dirac spinors may be written as
\bea
u(k_1) &=& \bigg({k_1 \cdot \sigma +m_1 \over \sqrt{2(k_1^0 +m_1)}}\xi_1 \ \ \ \ \ \
{k_1 \cdot \bar \sigma +m_1 \over \sqrt{2(k_1^0 +m_1)}}\xi_1 \bigg)^T\,,
\nonumber\\
v(k_2) &=& \bigg({k_2 \cdot \sigma +m_2 \over \sqrt{2(k_2^0 +m_2)}}\xi_2 \ \ \ \ \ \
-{k_2 \cdot \bar \sigma +m_2 \over \sqrt{2(k_2^0 +m_2)}}\xi_2 \bigg)^T\,,
\eea
where the particles have four-momenta,
\bea
k_1^\mu &=& (E_1, \overrightarrow{k}_1) = \left(\sqrt{m_1^2 + \overrightarrow{k}^2}, \overrightarrow{k} \right)
\nonumber\\
k_2^\mu &=& (E_2, \overrightarrow{k}_2) = \left(\sqrt{m_2^2 + \overrightarrow{k}^2}, -\overrightarrow{k} \right) .
\eea

The bilinears for an outgoing fermion/anti-fermion pair are then,
\bea
\bar u(k_1) v(k_2)
&=&
\left[\sqrt{E_1 + m_1 \over E_2 + m_2} + \sqrt{E_2 + m_2 \over E_1 + m_1} \right]
\overrightarrow{k} \cdot \left( \xi_1^\dagger \overrightarrow{\sigma} \xi_2 \right)\,,
\nonumber\\
\bar u(k_1) \gamma^5 v(k_2) &=&
 -{1\over \sqrt{(E_1 + m_1)(E_2 + m_2)}} \left[(E_1 + m_1)(E_2 + m_2) +\overrightarrow{k}^2  \right] \left( \xi_1^\dagger \xi_2 \right)\,,
\nonumber\\
\bar u(k_1) \gamma^0 \gamma^5 v(k_2)
&=&
-{1\over \sqrt{(E_1 + m_1)(E_2 + m_2)}}  \left[(E_1+m_1)(E_2+m_2) -\overrightarrow{k}^2 \right] \left(\xi_1^\dagger \xi_2 \right)\,,
\nonumber\\
\bar u(k_1) \gamma^i \gamma^5 v(k_2)
&=& \imath \epsilon^{ijk} k^j \left(\sqrt{E_2 + m_2 \over E_1 + m_1} + \sqrt{E_1 + m_1 \over E_2 + m_2} \right)
\left( \xi_1^\dagger \sigma^k \xi_2 \right)
+ k^i \left(\sqrt{E_1 + m_1 \over E_2 + m_2} - \sqrt{E_2 + m_2 \over E_1 + m_1} \right)  \left( \xi_1^\dagger \xi_2 \right)\,,
\nonumber\\
\bar u(k_1) \gamma^0 v(k_2)
&=& {E_1 +m_1 - E_2 - m_2 \over \sqrt{(E_1 + m_1)(E_2 + m_2)}} k^i  \left( \xi_1^\dagger  \sigma^i  \xi_2 \right)\,,
\nonumber\\
\bar u(k_1) \gamma^i v(k_2)
&=& - {(E_1 +m_1)(E_2 + m_2) +\overrightarrow{k}^2 \over \sqrt{(E_1 + m_1)(E_2 + m_2)}} \left( \xi_1^\dagger  \sigma^i \xi_2 \right)
+ {2k^i k^j \over \sqrt{(E_1 + m_1)(E_2 + m_2)}}  \left( \xi_1^\dagger \sigma^j \xi_2 \right)\,,
\nonumber\\
\bar u(k_1) \sigma^{0i} v(k_2)
&=& -\imath {(E_1 +m_1)(E_2 +m_2) -\overrightarrow{k}^2 \over \sqrt{(E_1 + m_1)(E_2 + m_2)} }  \left( \xi_1^\dagger  \sigma^i \xi_2 \right)
-2\imath {k^i k^j \over \sqrt{(E_1 + m_1)(E_2 + m_2)}}  \left( \xi_1^\dagger \sigma^j \xi_2 \right)\,,
\nonumber\\
\bar u(k_1) \sigma^{0i}\gamma^5 v(k_2)
&=& \imath  k^i \left(\sqrt{E_2 + m_2 \over E_1 + m_1} +\sqrt{E_1 + m_1 \over E_2 + m_2}  \right) \left( \xi_1^\dagger   \xi_2 \right)
+ \epsilon^{ijk}  k^j \left(\sqrt{E_2 + m_2 \over E_1 + m_1} -\sqrt{E_1 + m_1 \over E_2 + m_2}  \right)
\left( \xi_1^\dagger \sigma^k  \xi_2 \right)\,.
\eea

The bilinears of the initial state fermion/anti-fermion pair can be obtained by conjugating the above expressions.

In Table~\ref{ann} we provide the annihilation matrix elements for various dark matter
bilinears in the case of spin-0 and spin-1 dark matter.  For spin-1 dark matter,
the two particles have polarization vectors $\epsilon_1$ and $\epsilon_2$.

\begin{table}[t]
\begin{center}
\begin{tabular}{|c|c|}
  \hline
  bilinear &  annihilation matrix element\\
  \hline
  $\phi^\dagger \phi$ & 1 \\
  $\imath Im(\phi^\dagger \partial^0 \phi)$ & $0$ \\
  $\imath Im(\phi^\dagger \partial^i \phi)$ & $-\imath k^i$ \\
  $B_\mu B^\mu$ & $\epsilon_1 \cdot \epsilon_2$ \\
  $\imath Im(B^\dagger_\nu \partial^0 B^\nu)$ & $0$ \\
  $\imath Im(B^\dagger_\nu \partial^i B^\nu)$ & $- \imath k^i \epsilon_1^\dagger \cdot \epsilon_2$ \\
  $\imath (B^\dagger_i B_j - B^\dagger_j B_i)$ & $\imath (\epsilon^\dagger_{1i} \epsilon_{2j} - \epsilon^\dagger_{2i} \epsilon_{1j})$ \\
  $\imath (B^{i \dagger} B^0 - B^{0\dagger} B^i)$ & $\imath (\epsilon_1^{i\dagger} \epsilon_2^0 - \epsilon_2^{i\dagger} \epsilon_1^0 )$\\
  $\epsilon^{0ijk}B_i \partial_j B_k$ & $\imath \epsilon_{ijk} k^i (\epsilon_2^j \epsilon_1^k - \epsilon_2^k \epsilon_1^j) $ \\
  $\epsilon^{0ijk}B_j \partial_0 B_k$ & 0 \\
  $-\epsilon_{0ijk}B^0 \partial^j B^k$ & $\imath \epsilon_{ijk} k^j (\epsilon_2^0 \epsilon_1^k - \epsilon_2^k \epsilon_1^0) $ \\
  $B^\nu \partial_\nu B_0$ & $-2\imath E \epsilon_1^0 \epsilon_2^0- \imath k_i (\epsilon_2^i \epsilon_1^0 - \epsilon_2^0 \epsilon_1^i)$ \\
  $B^\nu \partial_\nu B^i$ & $-\imath E (\epsilon_2^0 \epsilon_1^i + \epsilon_1^0 \epsilon_2^i)- \imath k_j (\epsilon_2^j \epsilon_1^i - \epsilon_1^j \epsilon_2^i) $ \\
  \hline
\end{tabular}
\caption{The annihilation matrix elements for spin-0 and spin-1 dark matter bilinears.}
\label{ann}
\end{center}
\end{table}

One can verify that the structure $(\epsilon_1^0 \epsilon_2^k - \epsilon_1^k \epsilon_2^0)$ is only non-zero for an
initial state with total spin and $z$-axis spin projection $| S, S_z \rangle$ given by
$|2,1\rangle$, $|2,0 \rangle$, $| 2,-1 \rangle$ and $|0,0 \rangle$.  Similarly, the
structure $(\epsilon_1^j \epsilon_2^k - \epsilon_1^k \epsilon_2^j)$ is only non-zero for initial spin states
$|1,1\rangle$, $|1,0 \rangle$ and $|1,-1 \rangle$, and
the structure $(\epsilon_1^i \epsilon_2^0 + \epsilon_1^0 \epsilon_2^i)$ is only non-zero for initial spin
states $|1,1 \rangle$ and $|1,-1 \rangle$.
The structure $\epsilon_1 \cdot \epsilon_2$ is only non-zero for initial spin states
$|0,0 \rangle $ and $| 2,0 \rangle$.  Finally, $\epsilon_1^0 \epsilon_2^0$ is only non-zero
for initial spin states $|0,0 \rangle$ and $|2,0 \rangle$.




\end{document}